\documentclass{emulateapj}
\usepackage{graphicx}
\usepackage{txfonts}
\usepackage{natbib}
\usepackage[scaled]{helvet}
\usepackage{epsfig}
\usepackage{url}
\bibpunct{(}{)}{;}{a}{}{,}
\usepackage{color}
\usepackage[normalem]{ulem}

\begin{document}

\title{Recurring Occultations of RW Aurigae by Coagulated Dust in the Tidally Disrupted Circumstellar Disk}
\author{Joseph E. Rodriguez$^1$, Phillip A. Reed$^2$, Robert J. Siverd$^{3}$, Joshua Pepper$^{4}$, Keivan G. Stassun$^{1,5}$, B. Scott Gaudi$^6$, David A. Weintraub$^1$, Thomas G. Beatty$^{7,8}$, Michael B. Lund$^1$, and Daniel J. Stevens$^6$}

\affil{$^1$Department of Physics and Astronomy, Vanderbilt University, 6301 Stevenson Center, Nashville, TN 37235, USA}
\affil{$^2$Department of Physical Sciences, Kutztown University, Kutztown, PA 19530, USA}
\affil{$^3$Las Cumbres Observatory Global Telescope Network, 6740 Cortona Dr., Suite 102, Santa Barbara, CA 93117, USA}
\affil{$^4$Department of Physics, Lehigh University, 16 Memorial Drive East, Bethlehem, PA 18015, USA}
\affil{$^5$Department of Physics, Fisk University, 1000 17th Avenue North, Nashville, TN 37208, USA}
%\affil{$^6$Harvard-Smithsonian Center for Astrophysics, 60 Garden St, Cambridge, MA 02138, USA}
\affil{$^6$Department of Astronomy, The Ohio State University, Columbus, OH 43210, USA}
\affil{$^7$Department of Astronomy \& Astrophysics, The Pennsylvania State University, 525 Davey Lab, University Park, PA 16802}
\affil{$^8$Center for Exoplanets and Habitable Worlds, The Pennsylvania State University, 525 Davey Lab, University Park, PA 16802}
%\affil{$^{10}$Department of Physics and Astronomy, University of Rochester, Rochester, NY 14627-0171, USA}
%\affil{$^{11}$School of Physics, Astronomy and Computational Science, George Mason University, Fairfax, VA 22030, USA}
\shorttitle{Second Large Occultation of RW Aur A}

\begin{abstract}
	We present photometric observations of RW Aurigae, a Classical T Tauri system, that reveal two remarkable dimming events. These events are similar to that which we observed in 2010-2011, which was the first such deep dimming observed in RW Aur in a century's worth of photometric monitoring. We suggested the 2010-2011 dimming was the result of an occultation of the star by its tidally disrupted circumstellar disk. In 2012-2013, the RW Aur system dimmed by $\sim$0.7 mag for $\sim$40 days and in 2014/21015 the system dimmed by $\sim$2 mag for $>$250 days. The ingress/egress duration measurements of the more recent events agree well with those from the 2010-2011 event, providing strong evidence that the new dimmings are kinematically associated with the same occulting source as the 2010-2011 event. Therefore, we suggest that both the 2012-2013 and 2014-2015 dimming events, measured using data from the Kilodegree Extremely Little Telescope and the Kutztown University Observatory, are also occultations of RW Aur A by the tidally disrupted circumstellar material. Recent hydrodynamical simulations of the eccentric fly-by of RW Aur B suggest the occulting body to be a bridge of material connecting RW Aur A and B. These simulations also suggest the possibility of additional occultations which are supported by the observations presented in this work. The color evolution of the dimmings suggest that the tidally stripped disk material includes dust grains ranging in size from small grains at the leading edge, typical of star forming regions, to large grains, ices or pebbles producing grey or nearly grey extinction deeper within the occulting material. It is not known whether this material represents arrested planet building prior to the tidal disruption event, or perhaps accelerated planet building as a result of the disruption event, but in any case the evidence suggests the presence of advanced planet building material in the space between the two stars of the RW Aur system. 
\end{abstract}

\maketitle
\section{\bf{Introduction}}
The circumstellar environment of young stars of a few Myrs old (T Tauri stars) involves complex dynamical interactions between dust and gas that directly influences the formation of planets. Past studies have shown that binarity is a common property of T Tauri stars (Ghez et al. 1993, Leinert et al. 1993, Richichi et al. 1994, Simon et al. 1995, Ghez et al. 1997). The process of planetary formation can be significantly altered when the circumstellar disk is gravitationally influenced by a stellar companion \citep{Clarke:1993, Dai:2015}. Specifically, strong binary interactions with disks are also likely to influence planetary core formation and chemical composition by stirring up and heating materials, enhancing planetesimal collisions. The prototype example of this type of system is RW Aurigae, a binary system of two Classical T Tauri Stars (CTTS), RW Aur A and B \citep{Duchene:1999}. Detailed millimeter mapping by \citet{Cabrit:2006} showed evidence of a recent close stellar fly-by of RW Aur B which disrupted the circumstellar material around RW Aur A, leaving a short truncated circumstellar disk and a large $\sim$600 AU long tidal arm extending from RW Aur A. 

The system parameters are comprehensively described in \S2 of \citet{Rodriguez:2013}. Briefly, the RW Aur system is comprised of at least two components (RW Aur A and B) separated by $\sim$1.5\arcsec ($\sim$200 AU) \citep{Cabrit:2006}. \citet{Bisikalo:2012} measured that the separation of RW Aur A and B over $\sim$70 years has increased by $\sim$0.002\arcsec yr$^{-1}$. At the angular separation and 140 pc distance, the Keplerian orbital period would be $>$ 1500 years \citep{Bisikalo:2012}. It is likely that the orbit is likely coplaner, prograde, and either unbound or highly eccentric \citep{Dai:2015}. This suggests that the orbit of RW Aur A and B is inclined to our line-of-sight, like the disk around RW Aur A, at 45$^{\circ}$--60$^{\circ}$ \citep{Cabrit:2006}.

In 2010 the RW Aur system dimmed by $\sim$2 mags for a period of $\sim$180 days, marking the first event of this kind observed in this system dating back to the late 1890's \citep{Beck:2001}. \citet{Rodriguez:2013} (hereafter Paper I) interpreted that dimming as an occultation of RW Aur A by disrupted circumstellar material from the close fly-by encounter of the two stars RW Aur A and B. Using simple kinematic arguments, Paper I determined that the occulting object, likely a clump of circumstellar material, was $\sim$0.3 AU in width, moving at a few km/s. If in a Keplerian orbit, it would be $\sim$180 AU from RW Aur A . Using simple geometric and kinematic arguments, it was determined that the occulting feature could not be located in the circumstellar disk around RW Aur A and therefore may not be in Keplerian orbit. 

Recent hydrodynamical simulations by \citet{Dai:2015} support the interpretation that a star-disk tidal encounter during a fly-by of RW Aur B could explain the unusual morphology of the RW Aur system. They found a strong agreement between their simulations and the millimeter observations by \citet{Cabrit:2006}, which first proposed the star-disk fly-by scenario. The model predicts that the line of sight to RW Aur A currently intersects a bridge of stripped-off material between the two stars. \citet{Dai:2015} argue that the bridge structure may have small clumps of dense material that could occult the primary star. These simulations support the original hypothesis presented in Paper I and predict the possible occurrence of additional dimming events. In addition, numerical simulations of eccentric binary interactions of classical T Tauri stars suggest that the interaction can create accretion streams of inner disk material onto the stellar photosphere. These streams would be created near apastron and eventially form into ``ring-like'' structures around each star \citep{Sytov:2011,Gomez:2013}. 

In this paper, we present new high-cadence photometry of the RW Aur system showing a shallow dimming in 2012-2013 and a second, larger dimming event in 2014-2015. The 2014-2015 event was first reported by \citet{Petrov:2015} and resembles the dimming observed in late 2010 (Paper I). We apply similar geometric and kinematic arguments as we did for the 2010 event to show that the new dimmings are consistent with another clump of material from the tidally disrupted disk.

\section{\bf Photometric Observations}
Several photometric surveys have observed RW Aur over both short and long timescales going back to 1899. Here we describe the observations used in our analysis. 

\subsection{KELT-North}
Starting in 2003, the Kilodegree Extremely Little Telescope (KELT)-North survey has been continuously observing the entire sky between a declination of +18 and +44, searching for transiting Hot Jupiters around bright stars (8$<$$V$$<$10). Each KELT-North field spans $26^{\circ}$ $\times$ $26^{\circ}$ with 23\arcsec per pixel. All observations are in a broad $R$-band filter with a $\sim$15 min cadence \citep{Pepper:2007, Pepper:2012}. RW Aur is located in KELT-North Field 04, which is centered on $\alpha$ = 5hr 54m 14.466s, $\delta$ = $+31^{\circ}$ 44$\arcmin$ 37$\arcsec$ J2000. We obtained 9619 images of field 04 from UT 2006 October 27 to UT 2014 December 31. The data acquisition and reduction is described in detail in \S2 of \citet{Siverd:2012}. The KELT-North observations do not resolve individual stars in the RW Aur system.

\begin{table}
\centering
\caption{List of observations}
\begin{tabular}{c c c}
\hline
Filter & Exposure Time (s) & Number of Exposures \\
\hline
\textit{B} & 25 & 587 \\
\textit{V} & 15 & 602 \\
\textit{R} & 15 & 631 \\
\textit{I} & 15 & 604 \\
\hline
\end{tabular}
\label{table:observations}
\end{table}

\subsection{Kutztown University Observatory}
RW Aur was observed in \textit{BVRI} using the 0.61 m Ritchey-Chr\'{e}tien optical telescope at the Kutztown University Observatory (KUO) in Kutztown, Pennsylvania.  A total of 2305 data images were obtained, as listed in Table \ref{table:observations}, over 43 nights between UT 2014 February 25 and UT 2015 March 30.

\begin{figure}
%   \vspace{-1in}
\includegraphics[width=1\linewidth]{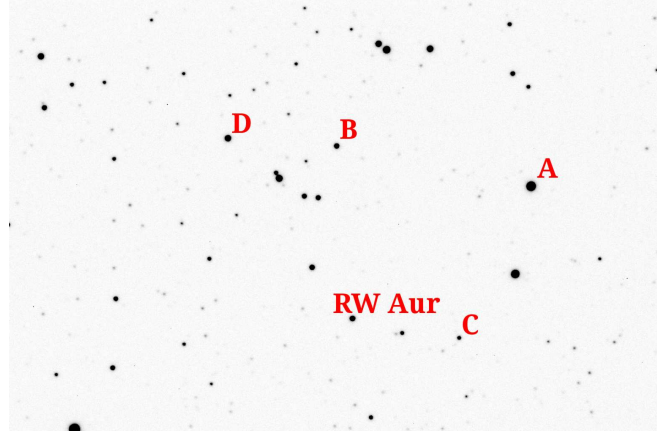}
%   \vspace{-1in}
\caption{The KUO field-of-view for RW Aur.  The standard reference stars are labeled as A, B, C, and D.}
\label{figure:FOV}
\end{figure}

The telescope's f/8 focal ratio, coupled with the camera's array of 3072 $\times$ 2048 (9 $\mu$m) pixels, yields a field of view of 19$\arcmin${.}5 $\times$ 13$\arcmin${.}0.  The CCD was kept at an operating temperature of -15$^{\circ}$C and dark, flat, and bias calibration frames were applied to all data images in the usual way.  A sample data image is given in Figure \ref{figure:FOV}, labeling RW Aur and four standard reference stars.  The known magnitudes of the reference stars are listed in Table \ref{table:reference_stars}.  A standard method of aperture photometry was employed, and the instrumental magnitudes were color-corrected using several Landolt standard fields. The KUO observations do not resolve the RW Aurigae system.

The observed \textit{BVRI} light curves are displayed in Figure \ref{figure:LC} and the {$B-V$} color curve is shown in Figure \ref{figure:Color_Curve}.

\begin{figure*}[!ht]
\centering\epsfig{file=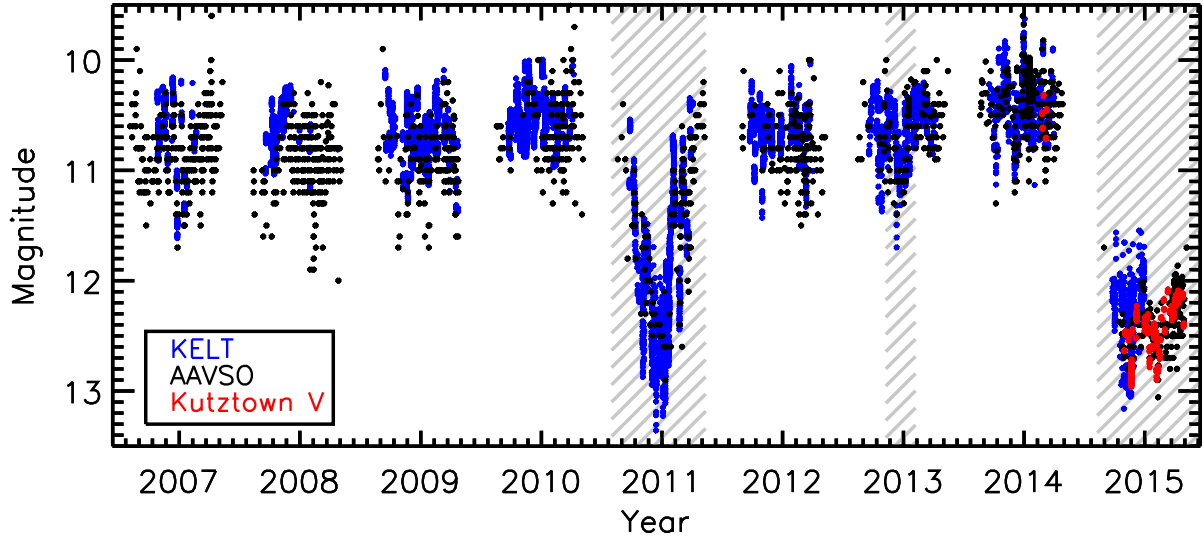,clip=,width=0.9\linewidth}
\caption{The KELT-North (Blue) and AAVSO (Black) observations plotted for the 9 KELT-North seasons. The three gray-shaded regions correspond to the 2010-2011, 2012-2013 and 2014-2015 large dimming events. The AAVSO and KUO data are in Visual and V-band magnitudes while the KELT-North observations are in instrumental magnitudes, that we approximate to the V-band but no attempt has been made to place all the data on the same absolute scale.}
\label{figure:KELTFull}
\end{figure*}

\begin{table*}
\centering
\caption{The properties of the reference stars. The quoted apparent magnitudes were obtained from the AAVSO Variable Star Database (A. A. Henden 2010, private communication). The cited sources are as follows: $^{\dagger}$APASS, $^{\dagger \dagger}$Tycho-2, $^{\dagger\dagger\dagger}$TASS.}
\begin{tabular}{l c c c c}
\hline \hline
  &   \multicolumn{3}{c}{Reference Star Label} \\\cline{1-5}
  & A & B & C & D \\
\hline
Tycho-ID &  TYC 2389-936-1 & TYC 2389-589-1 & N/A  & TYC 2389-630-1 \\
\hline
R.A. (J2000) & 05:07:24.62 & 05:07:50.55 & 05:07:35.27 & 5:08:05.17 \\
DEC. (J2000) & 30:20:28.1 & 30:19:02.5 & 30:24:47.0 & 30:18:40.6 \\
\textit{B} & 10.135 ($\pm$0.064)$^{\dagger\dagger}$ & 12.571 ($\pm$0.050)$^{\dagger}$ & 13.286 ($\pm$0.048)$^{\dagger}$  & 12.226 ($\pm$0.043)$^{\dagger}$ \\
\textit{V} & 9.617 ($\pm$0.040)$^{\dagger\dagger}$ & 12.046 ($\pm$0.012)$^{\dagger}$  & 12.901 ($\pm$0.017)$^{\dagger}$ & 11.437 ($\pm$0.011)$^{\dagger}$ \\
\textit{R} & --- & 11.665 ($\pm$0.072)$^{\dagger}$ & 12.582 ($\pm$0.069)$^{\dagger}$ & 10.980 ($\pm$0.050)$^{\dagger}$ \\
\textit{I} & 8.942 ($\pm$0.092)$^{\dagger\dagger\dagger}$ & 11.307 ($\pm$0.101)$^{\dagger}$ & 12.281 ($\pm$0.096)$^{\dagger}$ & 10.552 ($\pm$0.071)$^{\dagger}$ \\
($B-V$) & 0.518 ($\pm$0.075) & 0.525 ($\pm$0.051) & 0.385 ($\pm$0.051) & 0.789 ($\pm$0.044) \\
\hline
\end{tabular}
\label{table:reference_stars}
\end{table*}

\subsection{American Association of Variable Star Observers (AAVSO)}
AAVSO is a dedicated, non-profit orginization with the primary goal of understanding all types of variable stars. The archive consists of data from astronomers, both amateur and professional, around the world. Reported observations for RW Aur begin in 1937 and the data used in this work are either in the $V$ band or visual observations. The AAVSO observations do not resolve the RW Aur system.

\section{Results}
In this section we review the results from the 2010-2011 eclipse and present new observations from the KELT-North Survey showing two additional dimming events in 2012-2013 and the 2014-2015 dimming event first announced by \citet{Petrov:2015}.

\subsection{2010-2011 Dimming}
Photometric analysis of the RW Aur system by the KELT-North and AAVSO surveys showed that in late 2010 the RW Aur system dimmed from $V\sim10.4$ to $V\sim12$ for $\sim$180 days. Since RW Aur B is too faint ($V\sim13.7$) to affect the total brightness of the system ($V\sim10.4$), we assumed the entire dimming was caused by a decrease in flux from RW Aur A. This corresponded to an 86$\%$ reduction in the star's flux.  Spectroscopic observations by \citet{Chou:2013}, which coincided with the first half of the 2010-2011 dimming, suggest that the accretion behavior of RW Aur A was consistent with previous observations prior to the large dimming. This provides evidence that the dimming is independent of the close-in star-disk accretion process. Paper I modeled the dimming as an occultation of RW Aur A by a large body which possessed a sharp leading edge perpendicular to its direction of motion. Combining this model with kinematic and geometric arguments, Paper I argued that RW Aur A was occulted by a large ($\sim$0.3 AU wide) body moving at a maximum velocity of $\sim$2.6 km s$^{-1}$ and if in Keplerian orbit, would be located $\sim$180 AU from the star. Since the known short disk (57 AU, \citet{Cabrit:2006} around RW Aur A is quite inclined to our line of sight ($>$ 30$^{\circ}$), in Paper I we argued that the occulting body could not lie within the disk plane.  

\subsection{2012-2013 Small Dimming}
\begin{figure}[!ht]
\centering\epsfig{file=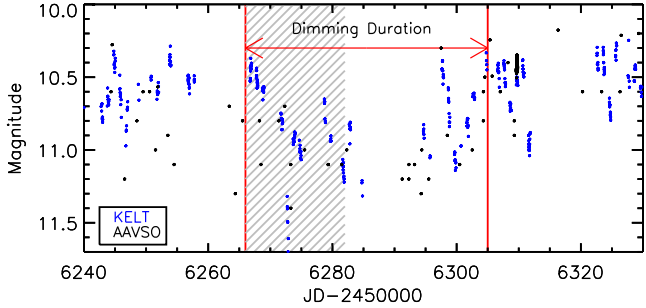,clip=,width=0.99\linewidth}
\caption{A zoom-in of Figure \ref{figure:KELTFull} in late 2012 to early 2013 showing a small dimming of the RW Aur system. The shaded region is the estimated ingress (the period of time for the dimming).}
\label{figure:SmallDimming}
\end{figure}

In the two seasons following the 2010-2011 dimming, the median brightness of the RW Aur system was slightly fainter then the median brightness of the season prior to the dimming ($V\sim10.5$). In early December of 2012, the brightness of the RW Aurigae system dimmed from a median brightness of $V = 10.5$ to $V = 11.2$ mag for $\sim$40 days (See Figure \ref{figure:SmallDimming}). As with the 2010 eclipse, if we assume the entire event is the result of only RW Aur A dimming, then RW Aur A dimmed by $\sim$50\%. Although our ability to estimate the ingress timescale is hindered by the known short-timescale photometric variability, we can constrain the ingress duration to be 10-20 days. We also estimate the egress to be similar in duration to the ingress. This event is about a third of the depth of the 2010-2011 event and significantly shorter in duration. However, the 10-20 day ingress timescale is similar to the ingress timescale of the 2010-2011 event. We adopt the same occultation model as in \S5.1 of Paper I, that is a large body with a sharp leading edge perpendicular to its direction of motion passing in front of RW Aur A. Since the ingress/egress timescale is similar to what was determined for the 2010-2011 event, this implies that the occulting bodies that caused the 2010-2011 and 2012-2013 dimmings are moving at a similar velocity and are located at a similar semi-major axis (that are 2.6 km s-1 and ~180 AU, respectively). The duration of the 2012-2013 event is only $\sim$40 days, which implies that the occulting body is 2.6 km s$^{-1}$ $\times$ 40 days = 0.06 AU in width (as compared to the $\sim$0.3 AU width estimated for the occulting body of the 2010-2011 event).  

\subsection{2014-2015 Dimming}
As first reported by \citet{Petrov:2015}, after the seasonal observing gap in mid-2014 the RW Aur system appeared significantly dimmer then in the previous observing season. From analyzing the H$\alpha$ and He I line at 5875\AA before and during the 2014 dimming, they find no evidence that the known high accretion rate of RW Aur A has changed during the time of the dimming. This suggests that the dimmings are unrelated to the accretion process which takes place close to the star. Their observations of the Na I D lines and the Ca II K line provide evidence that the stellar winds of RW Aur A have changed significantly and propose that the increased wind velocity is pushing dust from the disk across our line of sight. Resolved $UBVRI$ photometric observations of the RW Aur system during the 2014-2015 dimming indicated that RW Aur A was dimmer by $>$2 magnitudes in all bands and was actually fainter than its companion RW Aur B on UT 2014 November 14 (V$\sim$13.7) \citep{Antipin:2015}. X-ray observations during RW Aur's bright state and during the 2014-2015 dimming show that the absorbing column density increased during the dim state and was consistent with the interstellar medium's gas-to-dust ratio \citep{Schneider:2015}. Moreover, the resolved photometry by \citet{Antipin:2015} suggests that the dimming resulted from foreground grey extinction and also provides evidence that all three dimming events are the result of only RW Aur A becoming dimmer while the brightness of RW Aur B remained constant.

\begin{figure}[!ht]
\centering\epsfig{file=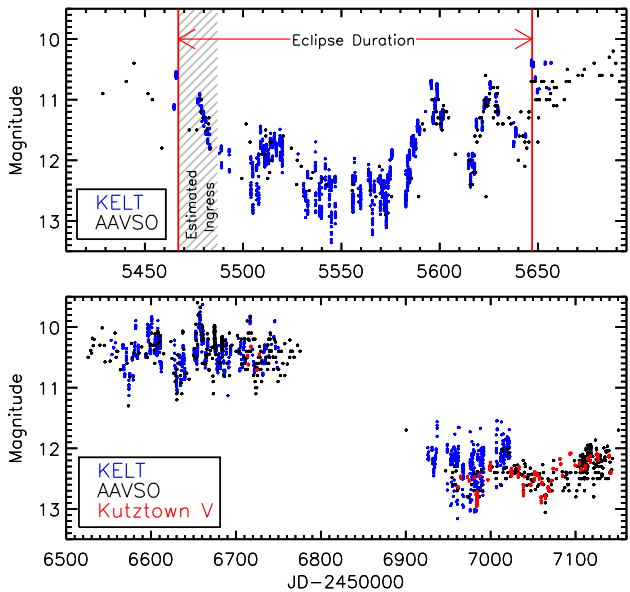,clip=,width=0.97\linewidth}
\caption{(Top) Recreation of Figure 3 from Paper I showing the 2010-2011 large dimming event. (Bottom) Zoom in of the last two KELT-North seasons showing the 2014-2015 dimming.}
\label{figure:KELTEclipse}
\end{figure}

Combining the KELT, AAVSO and KUO observations, we find that the combined RW Aur system dimmed by $\sim$2 mag in the KELT-North observations (Broad $R$ Band), similar in depth to the 2010-2011 dimming (See Figure \ref{figure:KELTFull}). Similar to the season prior to the 2010-2011 dimming, the RW Aur system was slightly brighter at a $V\sim10.4$ prior to the 2014-2015 dimming. Using KUO, we conducted multi-band ($BVRI$) photometric monitoring of the entire RW Aur system prior to and during the 2014-2015 dimming. We find that the depth is $\sim$2.3 mag in $B$, $\sim$2.0 mag in $V$, $\sim$1.75 mag in $R$ and $\sim$1.5 mag in $I$. \citet{Antipin:2015} showed that during the beginning of the 2014-2015 dimming, RW Aur A had dimmed by $\sim$3 mag and was actually fainter then RW Aur B in all bands in which we observed in ($BVRI$). Therefore, since our observations do not resolve the system, this depth difference is likely a result of the fact that the light from RW Aur B is included in our measurement. In \S4.2 we discuss that in fact it appears the dimming evolved with time, steadily becoming more grey.

Similar to the 2010-2011 dimming, the short-term non-periodic photometric variability that is so prominent outside the 2010-2011 dimming and prior to the 2014-2015 dimming is still apparent during the new event but has diminished significantly in amplitude. During the 2010-2011 dimming, a few large 1-1.5 mag brightening and re-dimming features were observed and attibruted to sub-structure in the occulting body (Paper I). In the AAVSO and KUO observations of the 2014-2015 dimming, we observe a $\sim$0.5 mag peak-to-peak amplitude brightening and then dimming event over an $\sim$80 day period beginning at JD-2450000 of $\sim$6970. Interestingly, this is similar in depth and duration to the entire event seen in 2012-2013. Observations at the end of the 2014-2015 observing season from KUO and AAVSO appear to suggest that the RW Aur system may have begun to return to its original median brightness. 

As with the previous two dimming events, we model the 2014-2015 dimming as an occultation of RW Aur A by a large body with a sharp leading edge perpendicular to its direction of motion. Unfortunately, the potential ingress and egress of this dimming appear to have occurred in the 2014 and 2015 seasonal observing gaps. Therefore, without an estimate of the ingress or egress timescale, we cannot calculate a transverse velocity of the occulting body. However, from the 2010-2011 and 2012-2013 dimming events, we calculated that the ingress timescale was between 10 and 30 days, which corresponds to a transverse velocity of 0.8-2.6 km s$^{-1}$. From the KELT-North, AAVSO, and KUO observations, the 2014-2015 dimming lasted at least the entire duration of the observing season, $\sim$250 days. Using this duration and adopting the calculated transverse velocity from the analysis of the 2010-2011 dimming suggests that the minimum width of the occulter is at least 2.6 km s$^{-1}$ $\times$ 250 days = 0.38 AU (0.8 km s$^{-1}$ $\times$ 250 days = 0.12 AU) since our total duration is only a lower estimate.

\begin{figure}[!ht]
\centering\epsfig{file=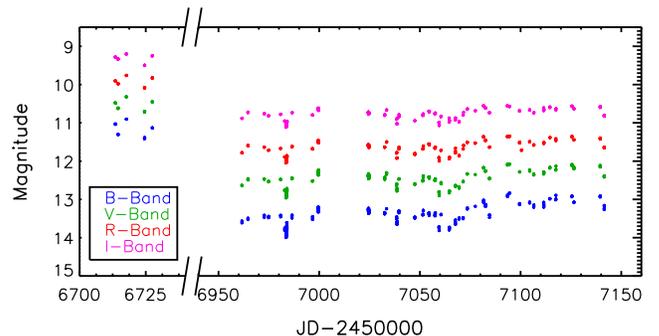,clip=,width=0.99\linewidth}
\caption{The KUO $BVRI$ light curves of RW Aur covering the 2014-2015 large dimming event. These observations do not resolve the RW Aur system.}
\label{figure:LC}
\end{figure}

\section{\bf Discussion}
In Paper I, we argued that the 2010-2011 dimming was caused by a consolidation of tidally disrupted material occulting RW Aur A. "In this section, we argue that the recent dimming events observed here, and first mentioned by \citet{Petrov:2015}, support this interpretation. We also discuss the possiblity of grain growth in the disrupted material.

\subsection{Interpretation: Occultation by the RW Aur A Tidally Disrupted Disk Material}

Using the IRAM Plateau de Bure Interferometer, the RW Aurigae system was mapped in $^{12}$CO and dust continuum \citep{Cabrit:2006}. Their observations showed a long tidal arm wrapped around RW Aur A. Through comparison with numerical simulations by \citet{Clarke:1993}, they proposed that RW Aur A recently experienced a fly-by from RW Aur B. This interaction would have significantly disrupted the disk originally around RW Aur A, resulting in the truncated disk and the large tidal arm. Also, the spectroscopic observations during the eclipse show that the accretion rate of RW Aur A did not change during the dimming \citep{Chou:2013}. Based on the millimeter and spectroscopic observations, Paper I proposed that the 2010-2011 dimming was caused by an occultation of the primary star, RW Aur A, by tidally disrupted material. This hypothesis has been supported by the hydrodynamical simulations by \citet{Dai:2015} which suggest the occulting body to be a bridge of disrupted material connecting RW Aur A and B. The simulations also predict the possibility of additional dimming events. 

Photometric monitoring of the RW Aur system from KELT-North, KUO and AAVSO show two additional dimming events that occurred after the 2010-2011 dimming event. From the 2012-2013 dimming, we estimate a similar ingress timescale as we did for the 2010-2011 event. We do not have coverage of ingress/egress for the 2014-2015 event due to the seasonal observing gaps. The similarities in the initial dimming duration for the 2010-2011 and 2012-2013 events suggest that the occulting bodies for both events are moving at similar velocities and likely at similar distances from RW Aur A (if we assume the occulting material to be in a Keplerian orbit). Therefore, it is likely that the cause of both (and possibly all three) dimming events are related. 

In this work and in Paper I, we made some simple assumptions (sharp leading edge and Keplerian motion) to determine some characteristics of the occulting bodies that caused the three dimming events observed. Our calculated transverse velocity from the 2010-2011 and 2012-2013 dimmings are consistent with the measured velocities of the tidally disrupted material from the millimeter observations by \citet{Cabrit:2006}. This velocity suggests a semi-major axis of $\sim$180 AU, which is less than the projected separation of RW Aur A and B ($\sim$200 AU) but larger then the estimated extent of the known disk around RW Aur A (57 AU, \citet{Cabrit:2006}). The hydrodynamical simulation by \citet{Dai:2015} of the RW Aur eccentric fly-by nicely replicates the millimeter observations by \citet{Cabrit:2006} and support the interpretation that the 2010-2011 occultation was caused by an occultation of RW Aur A by tidally disrupted material. The simulations also suggest the possibility of additional dimming events in the future. Our observations and analysis of the 2012-2013 and 2014-2015 dimmings are consistent with the simulations and interpretation first proposed in Paper I. 

Since no other dimming event was observed for $\sim$50 years prior to 2010 (Paper I) and two more dimmings have occurred since, it is probable that the 2010-2011 event was the leading front of tidally disrupted material and more dimming are likely to occur. From the hydrodynamical simulations, \citet{Dai:2015} suggested that the occulting mechanism may be a bridge of material connecting RW Aur A and B. Although this structure is poorly resolved in the simulations, they estimate it to be $\sim$100 AU wide ($\sim$0.7$\arcsec$ in the plane of the sky). If the 2010-2011 dimming was caused by the leading edge of this bridge of material, and is moving 0.8-2.6 km s$^{-1}$ as our calculations have shown, it will take 180-600 years for the trailing edge of the bridge to fully cross our line of sight. In $\sim$5 years, we have observed 3 separate dimming events with 3 different durations. Therefore, it is possible that $>$50 more dimmings of RW Aur could occur over the next century.

Young stellar objects have also been seen to eject blobs of gas with very high velocities. An example of this is HH 30, a young system in which gas blobs with sizes similar to our solar system being ejected from the star at $\sim$220 km s$^{-1}$ \citep{Burrows:1996}. These blobs are mesured to be $\sim$0.4$\arcsec$ in width (or $\sim$56 AU wide using the $\sim$140 parsec distance to the Taurus molecular cloud) \citep{Burrows:1996, Elias:1978}. If we assume that the cause of the RW Aur dimmings is caused by similarly sized gas blobs, the time required for one blob to cross our line of size at the observed velocity for the HH 30 gas blobs would be $\sim$1.2 years. This duration is similar to the duration of the 2010-2011 and 2014-2015 dimming events of RW Aur. Therefore, it is possible that the occulting body could be an ejected blob of gas from RW Aur A. This would also cause the spectroscopic signatures of gas in the line of sight that have been observed by \citet{Petrov:2015} for the 2014-2015 dimming.
However, we emphasize again that spectroscopic accretions signatures have been found to not correlate in time with the occultation events \citep{Chou:2013}, which suggests that the occulting material is not directly related to accretion phenomena which are generally believed to drive outflows and ejections.

Paper I investigated a series of other possible explanations for the 2010-2011 dimming event that are also ruled out for the additional two dimmings presented in this work. However, spectroscopic observations prior to and during the 2014-2015 dimming event show evidence that the stellar wind of RW Aur A has changed. Therefore, an alternative explanation for the 2014-2015 dimming is that enhanced stellar winds ejected large dust grains from the RW Aur A disk causing the occultation \citep{Petrov:2015}. We do not rule this interpretation out but suggest that the consistency between the millimeter observations of the disrupted material by \citet{Cabrit:2006}, our observed and calculated properties of the occulting bodies, and the agreement with the hydrodynamical simulations by \citet{Dai:2015} of the eccentric star-disk encounter, favor the tidally disrupted disk explantion as the cause of all three dimming events. Since there is no evidence of prior occultations for the last 50+ years (Paper I), and now we have several such events within the last few years, it is likely that we are observing the leading edge of the bridge structure, and the successive occultations presented here and in Paper I represent smaller coherent structures within it.

\begin{figure}[!ht]
\centering\epsfig{file=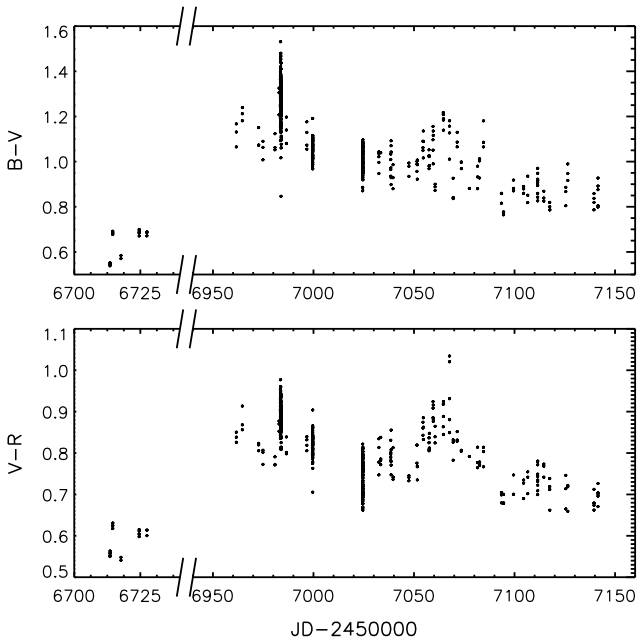,clip=,width=0.99\linewidth}
\caption{The KUO $B-V$(top) and $V-R$ (bottom) color curve of RW Aur A during the 2014-2015 dimming. The $BVR$ brightness of RW Aur B \citep{Antipin:2015} has been subtracted from the KUO observations }
\label{figure:Color_Curve}
\end{figure}

\subsection{Evidence for Grain Growth in the Tidally Disrupted Disk}

Although our color observations do not resolve the RW Aur system, observations by \citet{Antipin:2015} show that RW Aur B has remained constant during the 2014-2015 dimming, indicating that the color changes observed are likely not due to RW Aur B. Using the measure $BVR$ magnitudes for RW Aur B (14.5, 13.8 and 12.92 respectively, \citep{Antipin:2015}), we subtract the brightness of RW Aur B from the KUO $BVR$ observations to create the $B-V$ an $V-R$ color plots seen in Figure \ref{figure:Color_Curve}. The $B-V$ color of RW Aur increased from a quiescent value of $\sim0.6$ prior to the dimming, to $\sim$1.2 early in the 2014-2015 dimming, and then monotonically declined back toward the original value except for a brief reversal (JD-2450000 $\sim$ 7060) suggesting some inhomogeneity in the dust grain size distribution within the occulting material (see Figure \ref{figure:Color_Curve}). Meanwhile, the $V$ mag of the star during the middle of the dimming remained roughly constant, dimmed by $\sim$2.0 mag (see Figure \ref{figure:LC}). That the total extinction is roughly constant but the color of the extinction is changing indicates that the total absorption column is roughly constant throughout the occultation but that the ratio of total-to-selective extinction changes. The initial change in $B-V$ from 0.6 to 1.0 indicates a $B-V$ color excess, E($B-V$) = 0.4. With $A_V$ = 2.0, we have a ratio of total-to-selective extinction, $R(V) = A_V / E(B-V)$ = 2.0/0.4 = 5. This is the standard value typically adopted for dust grains in molecular clouds and star forming regions, which suggests the presence of dust grains, and not just molecular material, in the environment of this young system. The E($B-V$) value then drops steadily to $\sim$0.2 and perhaps less, giving $R(V)$ $>$ 2.0/0.2 $>$ 10. In other words, the occulting material becomes steadily more grey. Rather, the shift in observed $R(V)$ indicates that the dust grains at the leading edge of the occulting feature are relatively small and representative of dust in star forming regions, whereas the material deeper into the occulter is likely comprised of larger grains, which could be dust grains onto which ice mantles have developed, and/or larger coagulated grains or pebbles (assuming the extinction is optically thin). Another possibility is that the steadily greyer extinction is caused by having a larger fraction of the occulting material be optically thick as the dimming progressed. \citet{Antipin:2015} find that the RW Aur A spectral energy distribution changes during the 2014-2015 occultation and they suggest both a grey extinction and a selective extinction, similar to our results here. We do not know whether this evolved protoplanetary material existed prior to the tidal disruption event, or if the growth of the dust grains was aided by the disruption event. The evidence suggests that the building blocks of planetary material can exist in the space between binary stars, perhaps through fly-by interactions such as that seems to have occurred in the RW Aur system.

\section{\bf Summary and Conclusions}
With the deep dimming event observed in 2010-2011 (Paper I), the intrigue surrounding the RW Aur system has dramatically increased. New photometric observations from KELT, AAVSO and KUO show two additional dimming events in 2012-2013 and 2014-2015 (the 2014-2015 dimming was first announced by \citet{Petrov:2015}). From our analysis, the observations of the additional dimming events are consistent with an occultation of RW Aur A by tidally disrupted material lying far outside the extent and plane of its short circumstellar disk. This interpretation has been supported by hydrodynamical simulations of the proposed RW Aur star-disk interaction \citep{Dai:2015}. 

Multiband photometric observations during the 2014-2015 dimming showed that the occulting material became steadily more grey as the dimming progressed. This suggests that the outer portion of the occulting body consists of small dust grains while the core is primarily made up of (either or all three) large dust grains, dust enveloped in ice or is optically thick. Either way, the observations are consistent with evolved protoplanetary material and the evolution of this material may have been expedited by the tidal interaction. The recurring dimmings of RW Aur A will continue to increase the interest surrounding this unique system. Continued monitoring of RW Aur will provide insight into the effect the tidal interaction will have on the evolution of planetesimals from the disrupted material. Future high spatial resolution observations of this system would be of great value in clarifying the nature of the circumstellar environment.

\acknowledgments

Early work on KELT-North was supported by NASA Grant NNG04GO70G. J.A.P. and K.G.S. acknowledge support from the Vanderbilt Office of the Provost through the Vanderbilt Initiative in Data-intensive Astrophysics. This work has made use of NASA's Astrophysics Data System and the SIMBAD database operated at CDS, Strasbourg, France.

Work by B.S.G. and T.G.B. was partially supported by NSF CAREER Grant AST-1056524.

We acknowledge with thanks the variable star observations from the AAVSO International Database contributed by observers worldwide and used in this research.

\bibliographystyle{apj}

\bibliography{RW_AUR2}

\begin{thebibliography}{}
\expandafter\ifx\csname natexlab\endcsname\relax\def\natexlab#1{#1}\fi

\bibitem[{{Antipin} {et~al.}(2015){Antipin}, {Belinski}, {Cherepashchuk},
  {Cherjasov}, {Dodin}, {Gorbunov}, {Lamzin}, {Kornilov}, {Kornilov},
  {Potanin}, {Safonov}, {Senik}, {Shatsky}, \& {Voziakova}}]{Antipin:2015}
{Antipin}, S., {Belinski}, A., {Cherepashchuk}, A., {et~al.} 2015, Information
  Bulletin on Variable Stars, 6126, 1

\bibitem[{{Beck} \& {Simon}(2001)}]{Beck:2001}
{Beck}, T.~L., \& {Simon}, M. 2001, \aj, 122, 413

\bibitem[{{Bisikalo} {et~al.}(2012){Bisikalo}, {Dodin}, {Kaigorodov}, {Lamzin},
  {Malogolovets}, \& {Fateeva}}]{Bisikalo:2012}
{Bisikalo}, D.~V., {Dodin}, A.~V., {Kaigorodov}, P.~V., {et~al.} 2012,
  Astronomy Reports, 56, 686

\bibitem[{{Burrows} {et~al.}(1996){Burrows}, {Stapelfeldt}, {Watson}, {Krist},
  {Ballester}, {Clarke}, {Crisp}, {Gallagher}, {Griffiths}, {Hester},
  {Hoessel}, {Holtzman}, {Mould}, {Scowen}, {Trauger}, \&
  {Westphal}}]{Burrows:1996}
{Burrows}, C.~J., {Stapelfeldt}, K.~R., {Watson}, A.~M., {et~al.} 1996, \apj,
  473, 437

\bibitem[{{Cabrit} {et~al.}(2006){Cabrit}, {Pety}, {Pesenti}, \&
  {Dougados}}]{Cabrit:2006}
{Cabrit}, S., {Pety}, J., {Pesenti}, N., \& {Dougados}, C. 2006, \aap, 452, 897

\bibitem[{{Chou} {et~al.}(2013){Chou}, {Takami}, {Manset}, {Beck}, {Pyo},
  {Chen}, {Panwar}, {Karr}, {Shang}, \& {Liu}}]{Chou:2013}
{Chou}, M.-Y., {Takami}, M., {Manset}, N., {et~al.} 2013, \aj, 145, 108

\bibitem[{{Clarke} \& {Pringle}(1993)}]{Clarke:1993}
{Clarke}, C.~J., \& {Pringle}, J.~E. 1993, \mnras, 261, 190

\bibitem[{{Dai} {et~al.}(2015){Dai}, {Facchini}, {Clarke}, \&
  {Haworth}}]{Dai:2015}
{Dai}, F., {Facchini}, S., {Clarke}, C.~J., \& {Haworth}, T.~J. 2015, \mnras,
  449, 1996

\bibitem[{{Duch{\^e}ne} {et~al.}(1999){Duch{\^e}ne}, {Monin}, {Bouvier}, \&
  {M{\'e}nard}}]{Duchene:1999}
{Duch{\^e}ne}, G., {Monin}, J.-L., {Bouvier}, J., \& {M{\'e}nard}, F. 1999,
  \aap, 351, 954

\bibitem[{{Elias}(1978)}]{Elias:1978}
{Elias}, J.~H. 1978, \apj, 224, 857

\bibitem[{{G{\'o}mez de Castro} {et~al.}(2013){G{\'o}mez de Castro},
  {L{\'o}pez-Santiago}, {Talavera}, {Sytov}, \& {Bisikalo}}]{Gomez:2013}
{G{\'o}mez de Castro}, A.~I., {L{\'o}pez-Santiago}, J., {Talavera}, A.,
  {Sytov}, A.~Y., \& {Bisikalo}, D. 2013, \apj, 766, 62

\bibitem[{{Pepper} {et~al.}(2012){Pepper}, {Kuhn}, {Siverd}, {James}, \&
  {Stassun}}]{Pepper:2012}
{Pepper}, J., {Kuhn}, R.~B., {Siverd}, R., {James}, D., \& {Stassun}, K. 2012,
  \pasp, 124, 230

\bibitem[{{Pepper} {et~al.}(2007){Pepper}, {Pogge}, {DePoy}, {Marshall},
  {Stanek}, {Stutz}, {Poindexter}, {Siverd}, {O'Brien}, {Trueblood}, \&
  {Trueblood}}]{Pepper:2007}
{Pepper}, J., {Pogge}, R.~W., {DePoy}, D.~L., {et~al.} 2007, \pasp, 119, 923

\bibitem[{{Petrov} {et~al.}(2015){Petrov}, {Gahm}, {Djupvik}, {Babina},
  {Artemenko}, \& {Grankin}}]{Petrov:2015}
{Petrov}, P.~P., {Gahm}, G.~F., {Djupvik}, A.~A., {et~al.} 2015, \aap, 577, A73

\bibitem[{{Rodriguez} {et~al.}(2013){Rodriguez}, {Pepper}, {Stassun}, {Siverd},
  {Cargile}, {Beatty}, \& {Gaudi}}]{Rodriguez:2013}
{Rodriguez}, J.~E., {Pepper}, J., {Stassun}, K.~G., {et~al.} 2013, \aj, 146,
  112

\bibitem[{{Schneider} {et~al.}(2015){Schneider}, {G{\"u}nther}, {Robrade},
  {Facchini}, {Hodapp}, {Manara}, {Perdelwitz}, {Schmitt}, {Skinner}, \&
  {Wolk}}]{Schneider:2015}
{Schneider}, P.~C., {G{\"u}nther}, H.~M., {Robrade}, J., {et~al.} 2015, \aap,
  584, L9

\bibitem[{{Siverd} {et~al.}(2012){Siverd}, {Beatty}, {Pepper}, {Eastman},
  {Collins}, {Bieryla}, {Latham}, {Buchhave}, {Jensen}, {Crepp}, {Street},
  {Stassun}, {Gaudi}, {Berlind}, {Calkins}, {DePoy}, {Esquerdo}, {Fulton},
  {F{\H u}r{\'e}sz}, {Geary}, {Gould}, {Hebb}, {Kielkopf}, {Marshall}, {Pogge},
  {Stanek}, {Stefanik}, {Szentgyorgyi}, {Trueblood}, {Trueblood}, {Stutz}, \&
  {van Saders}}]{Siverd:2012}
{Siverd}, R.~J., {Beatty}, T.~G., {Pepper}, J., {et~al.} 2012, \apj, 761, 123

\bibitem[{{Sytov} {et~al.}(2011){Sytov}, {Kaigorodov}, {Fateeva}, \&
  {Bisikalo}}]{Sytov:2011}
{Sytov}, A.~Y., {Kaigorodov}, P.~V., {Fateeva}, A.~M., \& {Bisikalo}, D.~V.
  2011, Astronomy Reports, 55, 793

\end{thebibliography}

\end{document}